\documentclass[sn-mathphys-num]{sn-jnl}
\usepackage{tikz,fontawesome5}
\usepackage{pdfpages}
\usepackage{amsmath,amssymb,cancel}
\usepackage[normalem]{ulem}
\usepackage[T1]{fontenc}
\makeatletter
\gdef\orcidlogo{}%
\gdef\orcid#1{}%
\makeatother
\definecolor{alert}{RGB}{115,60,190}
\definecolor{question}{RGB}{35,145,95}

\DeclareMathOperator{\Var}{Var}
\DeclareMathOperator{\Cov}{Cov}

\newcommand{\avg}[1]{\langle #1\rangle}

\newcounter{assumption}

\title{Variance of the $SIS$ Epidemic on Networks:\newline A Diffusion Approximation}

\author[1]{\fnm{Lucija Nora} \sur{Farkaš}}
\author[1]{\fnm{Sebastian} \sur{Morel Balbi}}
\author[2]{\fnm{Hrvoje} \sur{Štefančić}}
\author[3,4]{\fnm{István Zoltán} \sur{Kiss}}
\author[1]{\fnm{Vinko} \sur{Zlatić}}

\affil[1]{\orgdiv{Theoretical Physics Division}, \orgname{Ruđer Bo\v{s}kovi\'{c} Institute}, \orgaddress{Bijeni\v{c}ka c. 54 10000 \city{Zagreb}, \country{Croatia}}}
\affil[2] {\orgname{Catholic University of Croatia, School of Medicine and Faculty of Engineering}, \orgaddress{ Ilica 244, 10000 \city{Zagreb}},\country{ Croatia}}
\affil[3]{\orgname{Network Science Institute, Northeastern University}, \orgaddress{\city{London}, \country{UK}}}
\affil[3]{\orgdiv{Department of Mathematics}, \orgname{Northeastern University}, \orgaddress{\city{Boston}, \state{MA}, \postcode{02115}, \country{USA}}}

\abstract{
Functional laws of large numbers (FLLNs) describe the mean-field trajectory of epidemics on networks, but say nothing about the fluctuations around it. These fluctuations are governed by moments of the degree distribution not relevant at the level of the mean. A rigorous functional central limit theorem (FCLT) exists for the susceptible--infected ($SI$) process on configuration-model graphs, but no analogue exists for $SIS$, where recovery reintroduces vertices into the susceptible pool with partially known neighborhoods, breaking the clean neighborhood distribution the $SI$ derivation relies on. We develop a tractable variance approximation for Markovian $SIS$ on configuration-model graphs, combining Gleeson's approximate master equation (AME) framework with a van Kampen system-size expansion in the spirit of the $SI$ FCLT. We derive a closed drift and diffusion matrix for a reduced susceptible/$SI$-edge/$SS$-edge count vector and obtain the time-dependent covariance via the associated Langevin/Lyapunov equation. Validation against Gillespie simulation across Poisson, regular, and power-law networks shows close agreement, with deviations near the epidemic threshold and in strongly heterogeneous networks.
}

\begin{document}

\maketitle

% ======================================================================
%  Introduction for the $SIS$-on-networks variance paper
%  (FCLT-style diffusion / Langevin approximation)
%
%  Requires a bibliography backend. With BibTeX:
%      \bibliographystyle{plain}      % or your journal's .bst
%      \bibliography{references}
%
%  Placeholder citation keys are defined in references.bib.
%  Adjust wording in the "contribution" paragraph to match what the
%  current drafts actually establish (3D aggregate vs. k-resolved, AME
%  comparison, etc.).
% ======================================================================

\section{Introduction}
\label{sec:intro}

Mathematical models of epidemics on networks are now a standard tool for
representing the heterogeneous contact structure through which infections
spread. The dominant analytical approach studies large-graph scaling
limits: as the number of vertices $N$ grows, suitably scaled compartment
counts converge to the solution of a system of ordinary differential
equations~(ODEs), a functional law of large numbers~(FLLN). Such limits
are now available for a range of dynamics on configuration-model graphs,
including the $SIR$ process~\cite{Decreusefond2012,Janson2014,Bohman2012,Barbour2013}
and multilayer variants~\cite{Jacobsen2018}.

The deterministic limit, however, describes only the mean behaviour. For
finite populations the realised trajectory fluctuates around this mean,
and for many practical purposes---quantifying forecast uncertainty,
estimating the probability of early extinction or of exceeding a
threshold prevalence, and designing control under risk---it is precisely
the \emph{size} of these fluctuations that matters. Crucially, the
variance of an epidemic on a network is not determined by the same
information as its mean. Graham and House~\cite{Graham2014} showed that,
while the mean prevalence in the early growth phase depends on the first
two moments of the degree distribution, the variance depends on the first
three, so that the skewness of the degree distribution drives the
stochastic spread of outcomes. Ball, Britton, Leung and
Sirl~\cite{Ball2019} made the point sharply: two configuration-model
ensembles with identical degree means can yield the same deterministic
limit yet different asymptotic variances. In short, network structure
exerts control over the variance of the process that is invisible at the
level of the mean, and any predictive framework that stops at the FLLN
discards this information.

Despite its importance, the variance has received far less rigorous
attention than the mean. The natural object is a functional central limit
theorem~(FCLT): a Gaussian-process description of the fluctuations of the
scaled counts around their FLLN limit, from which time-dependent
variances and covariances follow directly. For network epidemics,
closed-form variance results have largely been confined to the early
exponential-growth phase via branching-process or linearised diffusion
arguments~\cite{Graham2014,BallHouse2017}, or to final-size and
giant-component quantities~\cite{Ball2018,Ball2019,Barbour2019}. A
genuinely \emph{functional} CLT, valid over the body of the epidemic, was
obtained by KhudaBukhsh et al.~\cite{KhudaBukhsh2022} for the $SI$ process
on configuration-model graphs: the scaled vector of susceptible counts
and edge counts, recentred on the FLLN limit and rescaled by $\sqrt{N}$ (following the standard diffusion-approximation normalization in which the $O(N)$ count deviation from the LLN limit is blown up by $\sqrt{N}$ to yield an $O(1)$ Gaussian limit; see Eq.~(4.16) of \cite{KhudaBukhsh2022})
converges to a continuous Gaussian vector semimartingale, established
through a Doob--Meyer decomposition and Rebolledo's martingale
CLT~\cite{Rebolledo1980}. Their construction tracks not only compartment
sizes but also $SI$- and $SS$-edge counts, and shows that the limiting
covariance is governed by summary statistics of the degree distribution
through the generating-function operators $D^{2}\psi$ and $D^{3}\psi$%
---again, the second- and third-order degree information.

The $SI$ process is, however, the simplest network dynamic: infection is
irreversible and the susceptible neighborhood of a vertex can be
characterised by a single hypergeometric draw at the moment of infection.
Many diseases of interest instead admit reinfection, for which the $SIS$
model is the canonical recurrent-dynamics analogue. Extending an
FCLT-style variance calculation to $SIS$ is substantially harder: recovery
returns vertices to the susceptible pool with partially revealed
neighborhoods, so the clean neighborhood distribution that drives the
$SI$ analysis no longer applies directly, and the closures required to
express the drift and the diffusion matrix in terms of the aggregate
counts must be supplied approximately. Existing analyses of stochastic
$SIS$ dynamics on networks have largely focused on the endemic regime
through the quasi-stationary distribution~\cite{Overton2022,Cantwell2025}
rather than on the full temporal behaviour of the fluctuations.

In this paper we develop a tractable variance approximation for the
Markovian $SIS$ epidemic on configuration-model graphs in the spirit of the
$SI$ FCLT. We derive the deterministic drift and the diffusion (noise)
matrix for the susceptible-count and edge-count vector, using an approximate closure to describe the neighborhood of a susceptible node in a $SIS$ process analogous to the exact expression for the $SI$ case given by
KhudaBukhsh et al.~\cite{KhudaBukhsh2022}  and the
automated-moment-equation and pair-approximation framework of
Gleeson~\cite{Gleeson2013}, and we obtain the time-dependent covariance by
solving the associated Langevin/Lyapunov system. Because recovery breaks
the exact neighborhood characterisation, our $SIS$ construction is an
approximation based on a speculative idea that the  susceptible vertex’s neighborhood should
decorrelate from its infection history due to relatively common recoveries and reinfections rather than a proven FCLT; we therefore validate it against
direct stochastic simulation across a range of infection rates and degree
distributions,  and we examine where the higher-degree-moment closures and
the approximations to the structural operator $\kappa = D^{2}\psi$,  which we call $\kappa^{(2)}_S$ are
accurate and where they degrade. 
% --- Adapt the sentence above to your actual results (3D aggregate vs.
% --- k-resolved, AME comparison in Fig. X, etc.).

 The remainder of the paper is organized as follows.
Section~\ref{sec:model} sets up the model: Sec.~\ref{sec:stochastic$SIS$}
defines the stochastic $SIS$ process on an arbitrary graph, Sec.~\ref{sec:AME}
recapitulates the approximate master equation (AME) framework of
Gleeson~\cite{Gleeson2013}, which supplies the building blocks from which
we construct our observables, and Sec.~\ref{sec:VarianceEquations}
develops the Langevin/Lyapunov equations for the variance, in the spirit
of~\cite{KhudaBukhsh2022}, together with the assumptions and
approximations this requires. Section~\ref{sec:results} contains our
results: Sec.~\ref{sec:approx-pm-sk} closes the drift by approximating
the distribution of infected neighbors of a susceptible vertex,
Sec.~\ref{sec:nabla-kappa} treats the resulting susceptible-centered
wedge factor, and Sec.~\ref{sec:numerical} validates the approximation
numerically, showing that, despite lacking a formal proof of validity,
it performs well throughout the epidemic's duration across a range of
network types and epidemiological parameters. Section~\ref{sec: Discussion}
recapitulates the paper and lays out future directions of work.

\section{Model}
\label{sec:model}
In this section we first define the stochastic $SIS$ model on an arbitrary graph.

\subsection{Stochastic SIS on a graph}
\label{sec:stochastic$SIS$}

Let $G=(V,E)$ be an arbitrary finite graph with $N=|V|$ nodes and adjacency matrix $A$. The susceptible-infected-susceptible ($SIS$) process is formally defined as a continuous Markov process on the configuration space of dimension $2^N$. The state of each node $i$ is either $S\equiv 0$ if the node is susceptible or $I\equiv 1$ if the node is infected. We will denote the state of the node $i$ at time $t$ as $\chi_i(t)$. Infected nodes recover independently with the rate $\gamma$ i.e.
\begin{equation}
1 \longrightarrow 0
\qquad \text{with rate} \qquad
\gamma .
\end{equation}

Susceptible nodes can get infected by each of their infected neighbors with the rate $\beta$. More precisely, if node $i$ is susceptible, its instantaneous infection rate is proportional to the number of infected neighbors,
\begin{equation}
0 \longrightarrow 1
\qquad \text{with rate} \qquad
\beta \sum_{j=1}^N A_{ij}\chi_j(t),
\end{equation}
Therefore, for a configuration $\boldsymbol{\chi}=(\chi_1,\ldots,\chi_N)$, the transition rates can be written as 
\begin{equation}
t(\boldsymbol{\chi},\chi^{i,+})
=
\beta (1-\chi_i)\sum_{j=1}^N A_{ij}\chi_j,
\qquad
t(\boldsymbol{\chi},\chi^{i,-})
=
\gamma \chi_i ,
\end{equation}
where $\chi^{i,+}$ and $\chi^{i,-}$ denote the configurations obtained from $\boldsymbol{\chi}$ by setting $\chi_i=1$ iff $\chi_i=0$ before and $\chi_i=0$, iff it was $\chi_i=1$.

The total number of infected vertices is
\begin{equation}
N_I(t)=\sum_{i=1}^{N}\chi_i(t),
\end{equation}
and the corresponding infected fraction is
\begin{equation}
\rho(t)=\frac{N_I(t)}{N}.
\end{equation}
Similarly, the number of susceptible vertices is $N-N_I(t)$. Since recovered vertices immediately return to the susceptible class, the $SIS$ process has no removed states. The configuration in which all vertices are susceptible is absorbing, while all other configurations evolve through infection and recovery events. The total infection rate in the network is $\beta[SI]$, where $[SI]$ denotes the number of susceptible-infected edges, and the total recovery rate is $\gamma N_I$.

A particularly useful approach to deal with the dimensionality of the exact model is to use a mean-field approximation. Out of a number of mean field approximations that were developed to deal with stochastic $SIS$ models on graphs, we use the approximate master equation~\cite{Gleeson2013}, due to its versatility in dealing with heterogenous networks.

\subsection{Mean-field limit of the stochastic model: approximate master equation (AME)}
\label{sec:AME}
Let us recapitulate the AME framework devised by Gleeson~\cite{Gleeson2013}, let $S_{k,m}$, respectively $I_{k,m}$, be the set of vertices which are susceptible, and infected respectively, have degree $k$, and have $m$ infected neighbors. When the fractions of vertices in $S_{k,m}$ and $I_{k,m}$ are tracked, we say that the state of the system is neighborhood-resolved, while when only the fractions in $S_k$, the set of susceptible vertices that have degree $k$, are tracked, we say that it is degree-resolved. Note that we do not need to track the fractions in $I_k$, the set of infected vertices that have degree $k$, to have a degree-resolved state, as the expected numbers in $S_k$ and $I_k$ sets sum to $NP(k)$. In the $SIS$ process, susceptible vertices become infected through contacts with infected neighbors, while infected vertices recover and become susceptible again.

On these sets we define the conditional measures $\phi(S,m|k)$ and $\phi(I,m|k)$, which represent the fractions of degree-$k$ vertices that are susceptible or infected and have $m$ infected neighbors. 

Clearly,
\begin{align}
1
&=
\sum_{m=0}^{k}
\left[
\phi(S,m|k)+\phi(I,m|k)
\right].
\end{align}

It is also useful to define a fraction of infected nodes of degree $k$:
\begin{align}
\rho_k(t)
&\equiv
\sum_{m=0}^{k}\phi(I,m|k),
\\
1-\rho_k(t)
&=
\sum_{m=0}^{k}\phi(S,m|k).
\end{align}

The total fraction of infected vertices in the network is therefore
\begin{align}
\rho(t)
&=
\sum_k P(k)\rho_k(t)
\\
& =
\sum_k P(k)\sum_{m=0}^{k}\phi(I,m|k).
\end{align}
Consequently, the total number of infected vertices in a network containing $N$ vertices is
\begin{align}
N_I(t)
&=
N\rho(t).
\end{align}

Next, we recapitulate how the sizes of the $S_{k,m}$ and $I_{k,m}$ sets change in time. In the approximate master-equation framework, the evolution of the conditional probabilities is written as
\begin{align}
\frac{d}{dt}\phi(S,m|k)
&=
-F_{k,m}\phi(S,m|k)
+
R_{k,m}\phi(I,m|k)
\nonumber\\
&
+
\gamma_S
\left[
(m+1)\phi(S,m+1|k)
-
m\phi(S,m|k)
\right]
\nonumber\\
&
+
\beta_S
\left[
(\bar{m}+1)\phi(S,m-1|k)
-
\bar{m}\phi(S,m|k)
\right],
\label{eq:ame-sis-susceptible}
\end{align}

\begin{align}
\frac{d}{dt}\phi(I,m|k)
& =
F_{k,m}\phi(S,m|k)
-R_{k,m}\phi(I,m|k)
\nonumber\\
&
+
\gamma_I
\left[
(m+1)\phi(I,m+1|k)
-
m\phi(I,m|k)
\right]
\nonumber\\
&
+
\beta_I
\left[
(\bar{m}+1)\phi(I,m-1|k)-
\bar{m}\phi(I,m|k)
\right],
\label{eq:ame-sis-infected}
\end{align}

where $\bar{m}=k-m$. For the $SIS$ process, the node-level transition rates are
\begin{align}
F_{k,m}
&=
\beta m,
\nonumber\\
R_{k,m}
&=
\gamma .
\end{align}

Here $F_{k,m}$ is the infection rate of a susceptible degree-$k$ vertex with $m$ infected neighbors, while $R_{k,m}$ is the recovery rate of an infected vertex. The quantities $\beta_{S}$ and $\beta_{I}$ are the effective rates at which susceptible neighbors of susceptible and infected central vertices become infected, respectively. Similarly, $\gamma_{S}$ and $\gamma_{I}$ are the effective rates at which infected neighbors of susceptible and infected central vertices recover.
The AME closure gives the effective neighbor-infection rates as
\begin{align}
\beta_{S}
&=
\frac{
\sum_k P(k)\sum_{m=0}^{k}
\bar{m}F_{k,m}\phi(S,m|k)
}{
\sum_k P(k)\sum_{m=0}^{k}
\bar{m}\phi(S,m|k)
},
\nonumber\\
\beta_{I}
&=
\frac{
\sum_k P(k)\sum_{m=0}^{k}
mF_{k,m}\phi(S,m|k)
}{
\sum_k P(k)\sum_{m=0}^{k}
m\phi(S,m|k)
}.
\label{eq}
\end{align}
The effective neighbor recovery rates are:

\begin{align}
\gamma_{S}
&=
\frac{
\sum_k P(k)\sum_{m=0}^{k}
\bar{m}R_{k,m}\phi(I,m|k)
}{
\sum_k P(k)\sum_{m=0}^{k}
\bar{m}\phi(I,m|k)
},
\nonumber\\
\gamma_{I}
&=
\frac{
\sum_k P(k)\sum_{m=0}^{k}
mR_{k,m}\phi(I,m|k)
}{
\sum_k P(k)\sum_{m=0}^{k}
m\phi(I,m|k)
}.
\label{eq:effective recovery rates}
\end{align}

In Eqs.~\eqref{eq:ame-sis-susceptible} and \eqref{eq:ame-sis-infected}, terms with indices outside the allowed range $0\leq m\leq k$ are taken to be zero. For the $SIS$, $\gamma_I=\gamma_S=\gamma$.

Gleeson demonstrated that through these equations almost all binary processes on the networks can be traced even for very heterogenous networks, on the level of expected fractions. In what follows, we will demonstrate that the conditional probabilities of equations \ref{eq:ame-sis-susceptible} and \ref{eq:ame-sis-infected} enable effective computation of the variance of $SIS$ model for the broad class of networks. A note to the reader is also in place here, as AME has some limitations. For each degree $k$, one has to keep track of $k+1$ different neighborhoods described with parameter $m$. This means that for $k_{\min}\leqslant k \leqslant k_{\max}$, one has to track $(2+k_{\max}+k_{\min})(1+k_{\max}-k_{\min})$ differential equations. 
% --- Section map to be completed
\subsection{Variance equations}
\label{sec:VarianceEquations}
The approximate master equation described in the previous section gives a deterministic description of the stochastic $SIS$ process. In other words, it tells us how the expected fractions of vertices in the different classes evolve when the network is large. This deterministic description is the natural first approximation: for large $N$, the random epidemic trajectory is expected to remain close to its mean trajectory.

However, for any finite network, two realizations of the same $SIS$ process will not be identical. Even if they start from the same initial condition and use the same parameter values, the random order of infections and recoveries produces fluctuations around the deterministic AME solution. The goal of this section is to approximate the size of these fluctuations, and in particular to obtain an approximation for the time-dependent variance of the number of susceptible vertices.

We use a van Kampen~\cite{van1983stochastic} system-size expansion. If $\boldsymbol{X}(t)$ denotes a vector of stochastic counts, we write
\begin{align}
\boldsymbol{X}(t)
=
N\boldsymbol{\phi}(t)
+
\sqrt{N}\,\boldsymbol{\zeta}(t).
\end{align}
Here $\boldsymbol{\phi}(t)$ is the deterministic large-$N$ limit obtained from the AME, while $\boldsymbol{\zeta(t)}=(\boldsymbol{X}(t)-\langle\boldsymbol{X}(t)\rangle)/\sqrt{N}$ describes the leading-order fluctuations around that limit. Thus, the deterministic AME gives the centre of the distribution, and the fluctuation variable $\boldsymbol{\zeta}$ describes its spread.

The covariance matrix of these fluctuations is
\begin{align}
\boldsymbol{C}(t)
=
\Cov\left(\boldsymbol{\zeta}(t),\boldsymbol{\zeta}(t)^{\mathsf{T}}\right).
\end{align}
Once $\boldsymbol{C}(t)$ is known, the variance of the original count variables follows from the scaling above. For example, if the first component of $\boldsymbol{X}$ is the susceptible count $X_S=S$, then
\begin{align}\label{eq:Var_over_C}
\Var(X_S(t))
\approx
N\,C_{SS}(t).
\end{align}
Equivalently, the variance of the susceptible density $X_S/N$ is approximately $C_{SS}(t)/N$.

The covariance matrix satisfies a linear matrix differential equation of Langevin, or Lyapunov, type:
\begin{align}
\dot{\boldsymbol{C}}
=
\boldsymbol{J}\boldsymbol{C}
+
\boldsymbol{C}\boldsymbol{J}^{\mathsf{T}}
+
\boldsymbol{B}.
\label{eq:lyapunov-equation}
\end{align}
This equation has a simple interpretation. The first two terms describe how existing fluctuations are transported and stretched by the deterministic dynamics. This information is contained in the Jacobian matrix $\boldsymbol{J}$, which is obtained by differentiating the deterministic drift $\frac{\partial {\boldsymbol{\phi}}}{\partial t}$:
\begin{align}
\boldsymbol{J}(\boldsymbol{\phi})
=
\left.
\left(\nabla_{\tilde{\boldsymbol{\phi}}}\frac{\partial {\tilde{\boldsymbol{\phi}}}}{\partial t}
\right)\right|_{\tilde{\boldsymbol{\phi}}=\boldsymbol{\phi}}
\end{align}
The final term, $\boldsymbol{B}$, is the diffusion or noise matrix -- it describes the new randomness injected into the system by individual infection and recovery events. 

Thus, the construction of the variance approximation reduces to a bookkeeping problem: we must list the possible events, write down how each event changes the variables of interest, and weight these changes by their rates.\newline
To see which variables must be tracked, it is useful to first consider the variance of the susceptible count alone, as this quantity is what we want to find in the first place. For the $SIS$ process, infections occur at rate $\beta[SI]$, where $[SI]$ is the count of $SI$ edges, while recoveries occur at rate $\gamma I$. A direct jump-process calculation gives
\begin{align}
\frac{d}{dt}\Var(S)
=
\beta\langle[SI]\rangle
-
2\beta\Cov(S,[SI])
+
\gamma\langle I\rangle
-
2\gamma\Var(S).
\label{eq:dvars-jump}
\end{align}
This equation shows why the susceptible count cannot be treated in isolation. As usual, the variance of $S$ depends not only on the mean number of $SI$ edges $[SI]$, but also on higher moments, i.e. the covariance between $S$ and $[SI]$. Therefore, any closed approximation for $\Var(S)$ must also keep track of at least some information about the edge structure around susceptible vertices.\newline
In what follows we adapt the method of ~\cite{KhudaBukhsh2022} developed for the $SI$ model. We use AME computed variables to keep track of local correlations. 

A natural reduced choice is the three-dimensional count vector
\begin{align}
\boldsymbol{X}
=
\left(
X_S,\,
X_{SI},\,
X_{SS}
\right),
\end{align}
where
\begin{align}
X_S &= S, \qquad
X_{SI} = [SI]=\sum_{k=k_{\min}}^{k_{\max}}\sum_{m=0}^k m\,S_{k,m}, \qquad
X_{SS} = 2[SS]=\sum_{k=k_{\min}}^{k_{\max}}\sum_{m=0}^k\bar{m}\,S_{k,m}.
\end{align}
Here $X_{SS}$ counts susceptible--susceptible half-edges, so each $SS$ edge contributes twice. 
The deterministic scaled large-N limit of this count vector obtained from the AME is given by
\begin{align}
    \begin{split}
        \boldsymbol{\phi}=&\,(\phi(S),\,\phi([SI]),\,\phi([SS])):\\
        \phi(S)=&\,\sum_{k=k_{\min}}^{k_{\max}}\phi(S,k)=1-\rho(t),\qquad \phi(S,k)=P(k)\sum_{m=0}^k\phi(S,m|k),\\
        \phi([SI])=&\,\sum_{k=k_{\min}}^{k_{\max}}P(k)\sum_{m=0}^km\,\phi(S,m|k)=\dfrac{\langle [SI]\rangle}{N},\\
        \phi([SS])=&\,\sum_{k=k_{\min}}^{k_{\max}}P(k)\sum_{m=0}^k\bar{m}\,\phi(S,m|k)=\dfrac{2\langle [SS]\rangle}{N},
    \end{split}
\end{align}
where we have also introduced $\phi(S,k)$, the joint measure, which represents the fraction of vertices that are both susceptible and of degree $k$, since this measure will be useful in the following calculations.\newline
The remaining quantities can be recovered from conservation relations. In particular,
\begin{align}
\begin{split}
    X_I =&\, N-X_S, \qquad
X_{II} = E-2X_{SI}-X_{SS},\\
\phi(I)=&\,1-\phi(S),\qquad \phi([II])=\langle k\rangle-2\phi([SI])-\phi([SS]),
\end{split}
\end{align}
where $E=N \langle k\rangle$ denotes the total number of half-edges in the network and $\langle k\rangle$ is the expected degree of the configuration-model network, $\langle k\rangle=\sum_{k=k_{\min}}^{k_{\max}}k\,P(k)$. \newline
We now describe the jumps of this three-dimensional vector. Let
\begin{align}
X_{S,i}
=
\begin{cases}
1, & \text{if vertex } i \text{ is susceptible},\\
0, & \text{otherwise}.
\end{cases}
\end{align}
Let $m_i$ be the number of infected neighbors of vertex $i$. Then
\begin{align}
\bar{m}_i = k_i-m_i
\end{align}
gives the number of its susceptible neighbors. 
If a susceptible vertex becomes infected, then $X_S$ decreases by one. Its $SI$ edges become $II$ edges, while its $SS$ edges become $SI$ edges. Therefore the jump vector for an infection event is
\begin{align}
\Delta\boldsymbol{X}^{S\to I}_i
=
X_{S,i}\left(
-1,\,
\bar{m}_i-m_i,\,
-2\bar{m}_i
\right),
\qquad\text{with rate }
A^{S\to I}_i
=
\beta\, m_i.
\end{align}
Similarly, if an infected vertex recovers, then $X_S$ increases by one. Its $II$ edges become $SI$ edges, while its $SI$ edges become $SS$ edges. The recovery jump vector is
\begin{align}
\Delta\boldsymbol{X}^{I\to S}_i
=
(1-X_{S,i})\left(
+1,\,
m_i-\bar{m}_i,\,
2\bar{m_i}
\right),
\qquad\text{with rate }
A^{I\to S}_i
=
\gamma.
\end{align}
Because the jump vector itself contains the indicator $1-X_{S,i}$, only infected vertices contribute to this recovery term.\newline

Now that we have encoded the individual infection and recovery events into the jump vectors, we can write the explicit expression for the diffusion (noise) matrix:
\begin{align}
    \begin{split}
        \boldsymbol{B}=\boldsymbol{B}^{S\to I}+\boldsymbol{B}^{I\to S},\qquad \boldsymbol{B}^{\nu}=\dfrac{1}{N}\sum_{i=1}^NA^{\nu}_i\left(\Delta\boldsymbol{X}^{\nu}_i\right)\left(\Delta\boldsymbol{X}^{\nu}_i\right)^{\mathsf{T}},\qquad\nu\in\{S\to I,\,I\to S\},
    \end{split}
\end{align}
evaluated along the deterministic AME solution $\boldsymbol{X}=N\boldsymbol{\phi}$.\newline
Summing these jumps over all vertices gives the deterministic drift:
\begin{align}
\begin{split}
    \dot{X}_S
&=
-\beta X_{SI}
+
\gamma X_I,
\\
\dot{X}_{SI}
&=
\beta\left([SSI]-2[ISI]-X_{SI}\right)
+
\gamma\left(X_{II}-X_{SI}\right),
\\
\dot{X}_{SS}
&=
-2\beta[SSI]
+
2\gamma X_{SI}.
\end{split}
\label{eq:unclosed-drift}
\end{align}
The terms $[SSI]$ and $[ISI]$ are wedge counts. They are given by
\begin{align}
[SSI]
=
\sum_{k=k_{\min}}^{k_{\max}}
\sum_{m=0}^{k}
m\,\bar{m}\,X_{S_{k,m}},
\qquad
[ISI]
=
\sum_{k=k_{\min}}^{k_{\max}}
\sum_{m=0}^{k}
\binom{m}{2}X_{S_{k,m}}.
\end{align}
Here, $X_{S_{k,m}}$ is the count of susceptible, degree-$k$ vertices with $m$ infected neighbors, i.e., the number of vertices in the $S_{k,m}$ set.
The drift is therefore not yet closed in terms of only $X_S$, $X_{SI}$, and $X_{SS}$. For the closure of these wedge counts whose derivatives figure in the Jacobian, and later for other motifs arising in the calculation of the diffusion matrix $\mathbf{B}$, we still need an approximation for the number of infected neighbors around susceptible vertices, developed in Section \ref{sec:approx-pm-sk}, and a way to track the expected number of wedges centered on susceptible vertices,  developed in Section \ref{sec:nabla-kappa}.

\section{Results}
\label{sec:results}

\subsection{Approximating the distribution of infected neighbors}
\label{sec:approx-pm-sk}

 As a first step, we wish to assess the sums over $m$ by using
\begin{align}
X_{S_{k,m}}
\approx
P(m|S,k)\,X_{S_k},
\end{align}
and approximating $P(m|S,k)$, the probability that a vertex has $m$ infected neighbors conditioned on the vertex being susceptible and having degree $k$. Here $X_{S_k}$ is the number of susceptible vertices of degree $k$, i.e., the number of vertices in the $S_k$ set. In contrast, the reduced three-dimensional system tracks only $S$, $[SI]$, and $[SS]$. Therefore, in order to compute the Jacobian, we need to express all quantities as \emph{strict} functions of these three variables at least approximately, even though we can reconstruct these quantities from AME variables exactly without additional approximations. To close the reduced system, we approximate the number of infected neighbors of a susceptible degree-$k$ vertex by assuming that $P(m|S,k)$ behaves as if the neighbors of this vertex are sampled from the current pool of susceptible half-edges.
In this approximation, we get the hypergeometric expression
\begin{align}\label{eq: P(m|S,k)}
P(m|S,k)
\,\approx\,
\frac{
\binom{X_{SI}}{m}
\binom{X_{SS}}{\bar{m}}
}{
\binom{X_{SS}+X_{SI}}{k}
}.
\end{align}
This expression says that, conditional on a vertex being susceptible and having degree $k$, its $k$ incident half-edges are sampled from the pool consisting of $X_{SI}$ half-edges leading to infected neighbors and $X_{SS}$ half-edges leading to susceptible neighbors. While this is exact for an $SI$ process where the graph is dynamically constructed as infections percolate, as is shown by KhudaBukhsh et al.~\cite{KhudaBukhsh2022}, in the case of $SIS$ on a pre-set configuration-model graph, this is an approximation.

 In $SIS$ dynamics both the focal vertex and its neighbors repeatedly cycle between the susceptible
and infected states. We justify the approximation of the distribution of the neighborhood of a susceptible degree-$k$ vertex given by $P(m|S,k)$ in Eq. (\ref{eq: P(m|S,k)}) on the idea that each recovery event along an edge, and each subsequent reinfection, effectively re-randomizes the local correlation between a vertex’s current state and the states it has held in the past.
Provided several such renewal events occur along the relevant time scale, the composition of a susceptible
vertex’s neighborhood should decorrelate from its infection history and come to resemble a fresh draw from
the current global pool of half-edges, which is exactly the assumption underlying Eq. (\ref{eq: P(m|S,k)}). This argument
suggests two regimes in which the closure should be expected to degrade: (i) when recovery is slow relative
to infection ($\gamma\ll\beta$), so that few renewal events occur before the relevant correlations matter, and (ii) when
the graph is sparse or low-degree, so that a single neighbor’s state dominates the local environment and its
history-dependence cannot be averaged away by sampling from the aggregate pool. Both effects are
consistent with the discrepancies we observe numerically in Sec. \ref{sec:numerical} (see in particular the $k$ = 4 regular-graph
case).

The expression for $P(m|S,k)$ as given in Eq. (\ref{eq: P(m|S,k)}) allows us to close the sums over $m$ in the calculations of the wedge counts and the counts of other motifs that emerge in the calculation of the diffusion matrix. This brings us a step closer to being able to take a derivative of the deterministic drift needed for the calculation of the Jacobian in terms of solely $\phi(S)$, $\phi([SI])$ and $\phi([SS])$. 

This gives the wedge approximations
\begin{align}
\begin{split}
\dfrac{[SSI]}{N}
&\,\approx\dfrac{1}{N}\dfrac{X_{SI}\, X_{SS}}{(X_{SS}+X_{SI})_2}\sum_{k=k_{\min}}^{k_{\max}}(k)_2\,X_{S_k}\approx
\kappa^{(2)}_{S}
\frac{\phi\left([SI]\right)\,\phi\left([SS]\right)}{\phi\left(S\right)},\\
\dfrac{2[ISI]}{N}&\,\approx
\dfrac{1}{N}\dfrac{(X_{SI})_2}{(X_{SS}+X_{SI})_2}\sum_{k=k_{\min}}^{k_{\max}}(k)_2\,X_{S_k}\approx
\kappa^{(2)}_{S}
\frac{\phi\left([SI]\right)^{2}}{\phi\left(S\right)},
\end{split}
\end{align}
 where $(a)_b$ denotes the falling factorial
\begin{align}
    \begin{split}
(a)_b=\begin{cases}
    a(a-1)\cdots(a-b+1),&a\geqslant b;\\
    0,& a< b,
\end{cases}
    \end{split}
\end{align}
and we used the large-$N$ limit to switch to $\boldsymbol{\phi}$.
As $\kappa^{(2)}_S$ is related to  the expected number of susceptible-centered wedges we will refer to it as the susceptible-centered wedge factor in the following discussions. More generally, we define
\begin{align}\label{eq: mu over phiSk}
\kappa^{(r)}_S
=
\phi\left(S\right)^{r-1}
\frac{
\sum_k (k)_r\, \phi\left(S,k\right)
}{
\left(\sum_k k\,\phi\left(S,k\right)\right)^r
}=\left(\sum_{k=k_{\min}}^{k_{\max}}\phi\left(S,k\right)\right)^{r-1}\left[\frac{
\sum_k (k)_r\, \phi\left(S,k\right)
}{
\left(\sum_k k\,\phi\left(S,k\right)\right)^r
}\right],
\end{align}

Equivalently, since $\phi(S,k)/\phi(S)=P(k|S)$ and $(k)_r/r!$ is the number of ways to choose $r$ half-edges to form an $r$-star centered on a degree-$k$ vertex, we can interpret $\kappa^{(r)}_S$ as the normalized expected number of $r$-star motifs centered at a susceptible vertex, with normalization given by the $r^{\text{th}}$ power of the expected degree of a susceptble vertex and an ordering factor $r!$:
\begin{align}
\kappa^{(r)}_S
=
\frac{
\sum_k (k)_r\, P(k|S)
}{
\left(\sum_k k\,P(k|S)\right)^r
}=\dfrac{\langle (k)_r|S\rangle}{\langle k|S\rangle^r}=\dfrac{\left\langle \left.\displaystyle \binom{k}{r}\right|S\right\rangle}{\langle k|S\rangle^r/r!}.
\end{align} 
In the calculation of the diffusion (noise) matrix, we will also use the equivalently defined quantity for infected-centered motifs, $\kappa^{(r)}_I$:
\begin{align}\label{eq: muI over phiSk}
\begin{split}
    \kappa^{(r)}_I
=&\,
\phi\left(I\right)^{r-1}
\frac{
\sum_k (k)_r\, \phi\left(I,k\right)
}{
\left(\sum_k k\,\phi\left(I,k\right)\right)^r
}=\dfrac{\langle (k)_r|I\rangle}{\langle k|I\rangle^r},\\\phi(I)=&\,1-\phi(S),\qquad \phi(I,k)=P(k)-\phi(S,k)
,
\end{split}
\end{align}
where $\phi(I,k)$ stands for the joint measure which represents the fraction of vertices that are both infected
and of degree $k$.\newline
 Using these approximations, the reduced deterministic drift becomes
\begin{align}
\begin{split}
 \dot{\phi}_S
=&\,
-\beta\, \phi\left([SI]\right)
+
\gamma\, \phi_I,
\\
\dot{\phi}_{SI}
=&\,
\beta\,\kappa^{(2)}_S\,
\frac{
\phi\left([SI]\right)(\phi\left([SS]\right)-\phi\left([SI]\right))}{\phi\left(S\right)}
-
\beta\, \phi\left([SI]\right)
+
\gamma\left(\phi([II])-\phi\left([SI]\right)\right),
\\
\dot{\phi}_{SS}
=&\,
-2\,\beta\,\kappa^{(2)}_S
\frac{
\phi\left([SI]\right)\,\phi\left([SS]\right)}{\phi\left(S\right)}
+
2\,\gamma \, \phi\left([SI]\right).   
\end{split}
\label{eq:closed-drift}
\end{align}
which we use to construct the Jacobian matrix $\boldsymbol{J}$ in Eq.~\eqref{eq:lyapunov-equation}. The elements of the Jacobian are given in Table ~\ref{tab:Jacobian-matrix-elements}.

\begin{table}[t]
\centering
\begin{tabular}{c|c|c}
\hline
&&\\[-0.2em]
$J_{S,S} = \displaystyle -\gamma$
&
$J_{S,[SI]} = \displaystyle -\beta$
&
$J_{S,[SS]} = \displaystyle 0$
\\[0.8em]
\hline
\multicolumn{3}{c}{}\\[-0.2em]
\multicolumn{3}{c}{$J_{[SI],S} = \displaystyle -\beta\,\kappa^{(2)}_S\,\dfrac{\phi([SI])(\phi([SS])-\phi([SI]))}{\phi^2_S}$} \\[1.2em]
\multicolumn{3}{c}{$J_{[SI],[SI]} = \displaystyle \beta\,\kappa^{(2)}_S\,\dfrac{\phi([SS])-2\phi([SI])}{\phi(S)}-\beta-3\,\gamma$} \\[1.2em]
\multicolumn{3}{c}{$J_{[SI],[SS]} = \displaystyle \beta\,\kappa^{(2)}_S\,\dfrac{\phi([SI])}{\phi(S)}-\gamma$} \\[1.2em]
\hline
\multicolumn{3}{c}{}\\[-0.2em]
\multicolumn{3}{c}{$J_{[SS],S} = \displaystyle 2\,\beta\,\kappa^{(2)}_S\,\dfrac{\phi([SI])\,\phi([SS])}{\phi^2_S}$} \\[1.2em]
\multicolumn{3}{c}{$J_{[SS],[SI]} = \displaystyle -2\,\beta\,\kappa^{(2)}_S\,\dfrac{\phi([SS])}{\phi(S)}+2\,\gamma$} \\[1.2em]
\multicolumn{3}{c}{$J_{[SS],[SS]} = \displaystyle -2\,\beta\,\kappa^{(2)}_S\,\dfrac{\phi([SI])}{\phi(S)}$} \\[1.2em]
\hline
\end{tabular}
\caption{Matrix elements of the Jacobian $\boldsymbol{J}$, $J_{a,b}=\partial\phi(a)/\partial\phi(b)$.}
\label{tab:Jacobian-matrix-elements}
\end{table}

The same jump vectors are used to construct the diffusion matrix $\boldsymbol{B}$. The difference is that, for the drift, we sum rates $A^{\nu}_i$ times jumps $\Delta\boldsymbol{X}^{\nu}_i$ for $\nu\in\{S\to I,\,I\to S\}$, whereas for the diffusion matrix we sum products of rates and the outer products of jumps, resulting in the need to account for 3-stars, giving rise to $\kappa^{(3)}_S$ factors. The explicit forms of the infection and recovery contributions, $\boldsymbol{B}^{S\to I}$ and $\boldsymbol{B}^{I\to S}$, are given in Tables \ref{tab:infection-diffusion-matrix-elements} and \ref{tab:recovery-diffusion-matrix-elements}, respectively.

\begin{table}[t]
\centering
\begin{tabular}{c}
\hline
\\[-0.2em]
$B^{S\to I}_{S,S} = \displaystyle \beta\,\phi([SI])$
\\[1.2em]
$B^{S\to I}_{S,[SI]} = \displaystyle -\beta\,\kappa^{(2)}_S\dfrac{\phi([SI])(\phi([SS])-\phi([SI]))}{\phi(S)}
+
\beta \,\phi([SI])$
\\[1.2em]
$B^{S\to I}_{S,[SS]} = \displaystyle 2\,\beta\,\kappa^{(2)}_S\dfrac{\phi([SI])\,\phi([SS])}{\phi(S)}$
\\[1.2em]
\hline
\\[-0.2em]
$B^{S\to I}_{[SI],[SI]} = \displaystyle \beta\, \kappa^{(3)}_S\dfrac{\phi([SI])(\phi([SS])-\phi([SI]))^2}{\phi(S)^2}-
\beta\,\kappa^{(2)}_S\dfrac{\phi([SI])(\phi([SS])-3\phi([SI]))}{\phi(S)}
+
\beta\,\phi([SI])$ \\[1.2em]
$B^{S\to I}_{[SI],[SS]} = \displaystyle -2\,\beta \,\kappa^{(3)}_S
\dfrac{\phi([SI])\,\phi([SS])(\phi([SS])-\phi([SI]))}{\phi(S)^2}$ \\[1.2em]
\hline
\\[-0.2em]
$B^{S\to I}_{[SS],[SS]} = 4\,\beta \dfrac{\phi([SI])\,\phi([SS])}{\phi(S)}
\left(
\kappa^{(3)}_S\dfrac{\phi([SS])}{\phi(S)}
+\kappa^{(2)}_S
\right)\displaystyle $ \\[1.2em]
\hline
\end{tabular}
\caption{Matrix elements of the symmetric infection diffusion matrix $\boldsymbol{B}^{S\to I}$.}
\label{tab:infection-diffusion-matrix-elements}
\end{table}

\begin{table}[t]
\centering
\begin{tabular}{c|c|c}
\hline
&&\\[-0.2em]
$B^{I\to S}_{S,S} = \displaystyle \gamma\, \phi_I$
&
$B^{I\to S}_{S,[SI]} = \displaystyle \gamma\, (\phi([II])-\phi([SI]))$
&
$B^{I\to S}_{S,[SS]} = \displaystyle 2\,\gamma\,\phi([SI])$
\\[0.8em]
\hline
\multicolumn{3}{c}{}\\[-0.2em]
\multicolumn{3}{c}{$B^{I\to S}_{[SI],[SI]} = \displaystyle \gamma\,\kappa^{(2)}_I\,  \dfrac{\left(\phi([II])-\phi([SI])\right)^2}{\phi_I}
            
            +\gamma\,(\phi([II])+\phi([SI]))$} \\[1.2em]
\multicolumn{3}{c}{$B^{I\to S}_{[SI],[SS]} = \displaystyle 2\,\gamma \,\kappa^{(2)}_I\,\dfrac{\phi([SI])\left(\phi([II])-\phi([SI])\right)}{\phi_I}
            
            -2\,\gamma\,\phi([SI])$} \\[1.2em]
\hline
\multicolumn{3}{c}{}\\[-0.2em]
\multicolumn{3}{c}{$B^{I\to S}_{[SS],[SS]} = \displaystyle 4\,\gamma \,\phi([SI])
            \left(1+
                \kappa^{(2)}_I\dfrac{\phi([SI])}{\phi_I}
                
            \right)$} \\[1.2em]
\hline
\end{tabular}
\caption{Matrix elements of the symmetric recovery diffusion matrix $\boldsymbol{B}^{I\to S}$.}
\label{tab:recovery-diffusion-matrix-elements}
\end{table}

\subsection{Treatment of the susceptible-centered wedge factor}
\label{sec:nabla-kappa}

The reduced three-dimensional system depends on the quantity $\kappa^{(2)}_S$. This quantity is not determined by $\phi\left(S\right)$, $\phi\left([SI]\right)$, and $\phi\left([SS]\right)$ alone; it depends on the degree distribution of the susceptible population. In our implementation, we therefore compute $\kappa^{(2)}_S$ from the neighborhood-resolved AME solution and use it as an input to the reduced covariance system.

 However, the derivative of the drift, i.e., the Jacobian, should include derivatives of $\kappa^{(2)}_S$ with respect to the reduced variables $\phi(S)$, $\phi([SI])$ and $\phi([SS])$, but as $\kappa^{(2)}_S$ is a function of the $\{\phi\left(S,k\right)\}$ as evident from Eq. (\ref{eq: mu over phiSk}), these derivatives are not available.

To assess the importance of this term, we consider the $SI$ case by setting $\gamma=0$, where the corresponding susceptible degree distribution has an explicit form:
\begin{align}
P(k|S)
=
\frac{P(k)\vartheta^k}{\psi(\vartheta)},
\qquad
\psi(\vartheta)
=
\sum_k P(k)\vartheta^k.
\end{align}
where $\vartheta$ denotes the probability that a given half-edge attached to vertex $i$ has not transmitted infection to $i$ by time $t$ under the condition that $i$ started of susceptible at $t=0$.
In that setting, the deterministic limit of the susceptible fraction, $\phi\left(S\right)$, is related to $\vartheta$ by
\begin{align}
\phi\left(S\right)(t)
=
\phi\left(S\right)(0)\,\psi(\vartheta(t)).
\end{align}
Since the probability generating function $\psi$ is monotone, $\vartheta$ can be viewed as a function of $\phi\left(S\right)$, and hence $\kappa^{(2)}_S$ can also be viewed as a function of $\phi\left(S\right)$.

\begin{figure}[!ht]
    \centering
    \includegraphics[width=1.0\textwidth]{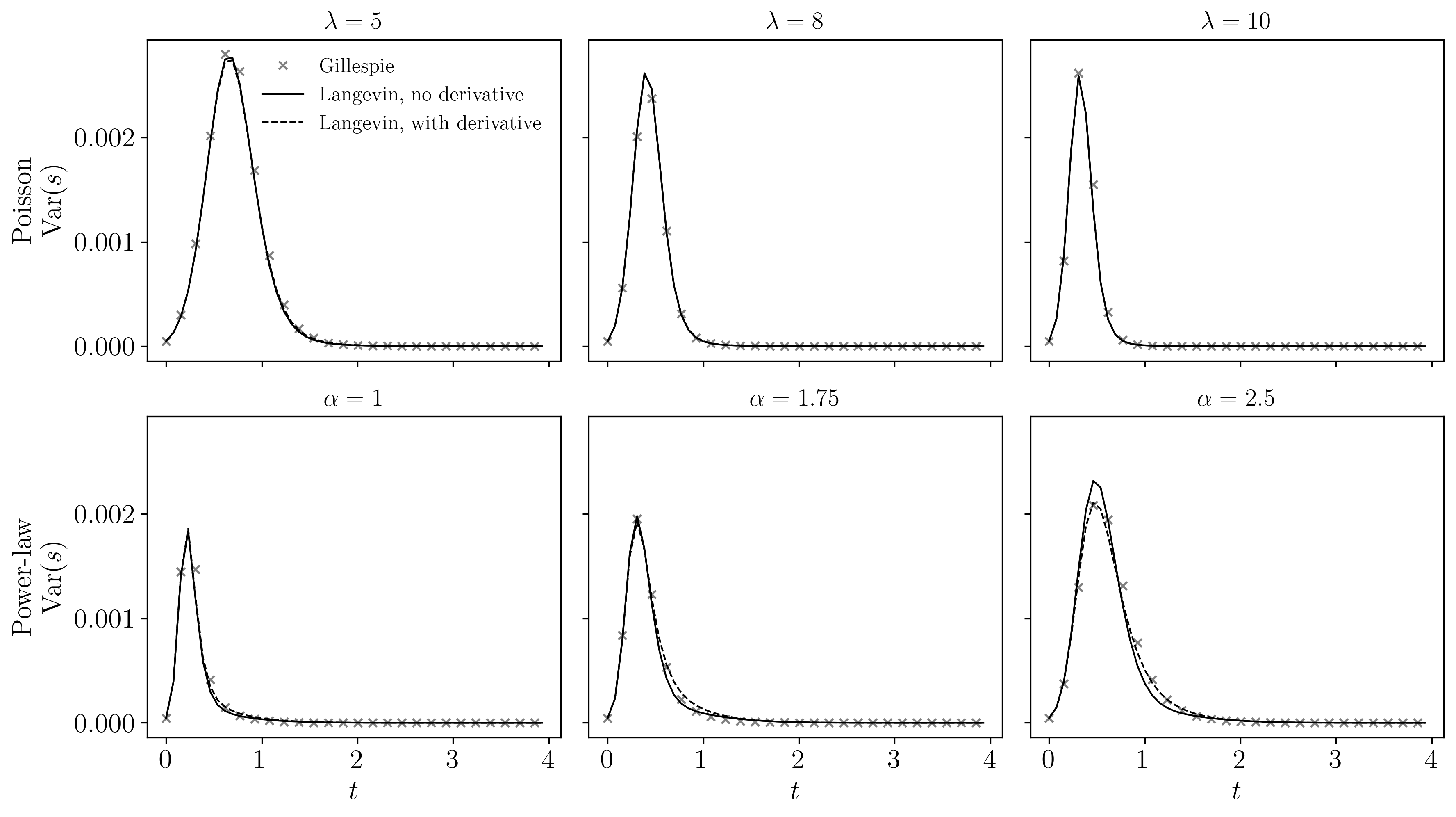}
    \caption{ Comparison of the AME-Langevin  prediction for the variance of the fraction of susceptibles, $s=X_S/N$, for $SI$ dynamics ($\gamma =0$), calculated with and without the $\partial \kappa^{(2)}_S/\partial\phi(S)$ term in the Jacobian, against the empirical variance from
Gillespie simulations. The two theoretical predictions are visually very similar, with the version including the derivative showing a marginal improvement in agreement with the simulation data. Data are measured across $N_r=1000$ Gillespie realizations of the $SI$ process on a configuration-model network with $N=1000$ vertices and either a Poisson degree distribution with parameter $\lambda=5,\, 8$ and $10$ or a power-law degree distribution  $P(k)\propto k^{-\alpha}$, $\alpha = 1,\,1.75$ and $2.5$. Both distributions are truncated so that $3\leqslant k\leqslant 30$.}
    \label{fig:SI_derivative}
\end{figure}

This gives, for $SI$ only,
\begin{align}
\frac{\partial \kappa^{(2)}_S}{\partial \phi\left(S\right)}
=
\frac{1}{\phi\left(S\right)}
\left[
\kappa^{(2)}_S
\left(
1-2\kappa^{(2)}_S
\right)
+
\kappa^{(3)}_S
\right],
\qquad
\frac{\partial \kappa^{(2)}_S}{\partial \phi\left([SI]\right)}
=
\frac{\partial \kappa^{(2)}_S}{\partial \phi\left([SS]\right)}
=
0.\label{eq:SecondApprox}
\end{align}
We then compare the variance obtained from Eq.~\eqref{eq:lyapunov-equation} with and without this derivative contribution. In the $SI$ test case,  the improvement when the derivative is included is negligible, as can be inferred from Fig. \ref{fig:SI_derivative}, where it is demonstrated on several Poisson and scale-free (power-law) graphs.

Motivated by this observation, we neglect the derivative of $\kappa^{(2)}_S$ in the $SIS$ covariance approximation and treat $\kappa^{(2)}_S$ as a time-dependent coefficient supplied by the AME solution. We note that this is exact for all processes on $k$-regular graphs, because all the nodes have the same degree. Therefore, $\kappa_S^{(2)}$ can not vary with the number of susceptible nodes i.e. $\frac{\partial \kappa^{(2)}_S}{\partial \phi\left(S\right)}$ of Eq.~\ref{eq:SecondApprox} is exactly zero and the susceptible-centered wedge factor is a constant, $\kappa^{(2)}_S=1-k^{-1}$. In contrast, we expect this approximation to be less reliable for strongly heterogeneous networks, as can be discerned from the degree-resolved derivative valid for both $SI$ and $SIS$ processes:
\begin{align}
    \begin{split}
        \dfrac{\partial\kappa^{(2)}_S}{\partial\phi(S,k)}=&\,\dfrac{\kappa^{(2)}_S}{\phi(S)}\left(1+\dfrac{k(k-1)}{\langle k(k-1)|S\rangle}-2\dfrac{k}{\langle k|S\rangle}\right).
    \end{split}
\end{align}

The final variance approximation is therefore obtained as follows. First, solve the AME system to obtain the deterministic trajectory and the degree-resolved quantities needed to evaluate $\kappa^{(2)}_S$, $\kappa^{(2)}_I$ and $\kappa^{(3)}_S$. Second, evaluate the Jacobian $\boldsymbol{J}$ while setting the derivatives of $\kappa^{(2)}_S$ to 0, and the diffusion matrix $\boldsymbol{B}$ along this trajectory. Finally, solve the Lyapunov equation
\begin{align}
\dot{\boldsymbol{C}}
=
\boldsymbol{J}\boldsymbol{C}
+
\boldsymbol{C}\boldsymbol{J}^{\mathsf{T}}
+
\boldsymbol{B}
\end{align}
to obtain the covariance matrix of the leading-order fluctuations. The desired variance of the susceptible count is then read from the first diagonal entry:
\begin{align}
\Var(X_S(t))
\approx
N C_{SS}(t).
\end{align}

\subsection{Numerical validation} \label{sec:numerical}

\begin{figure}[!ht]
    \centering
    \includegraphics[width=1.0\textwidth]{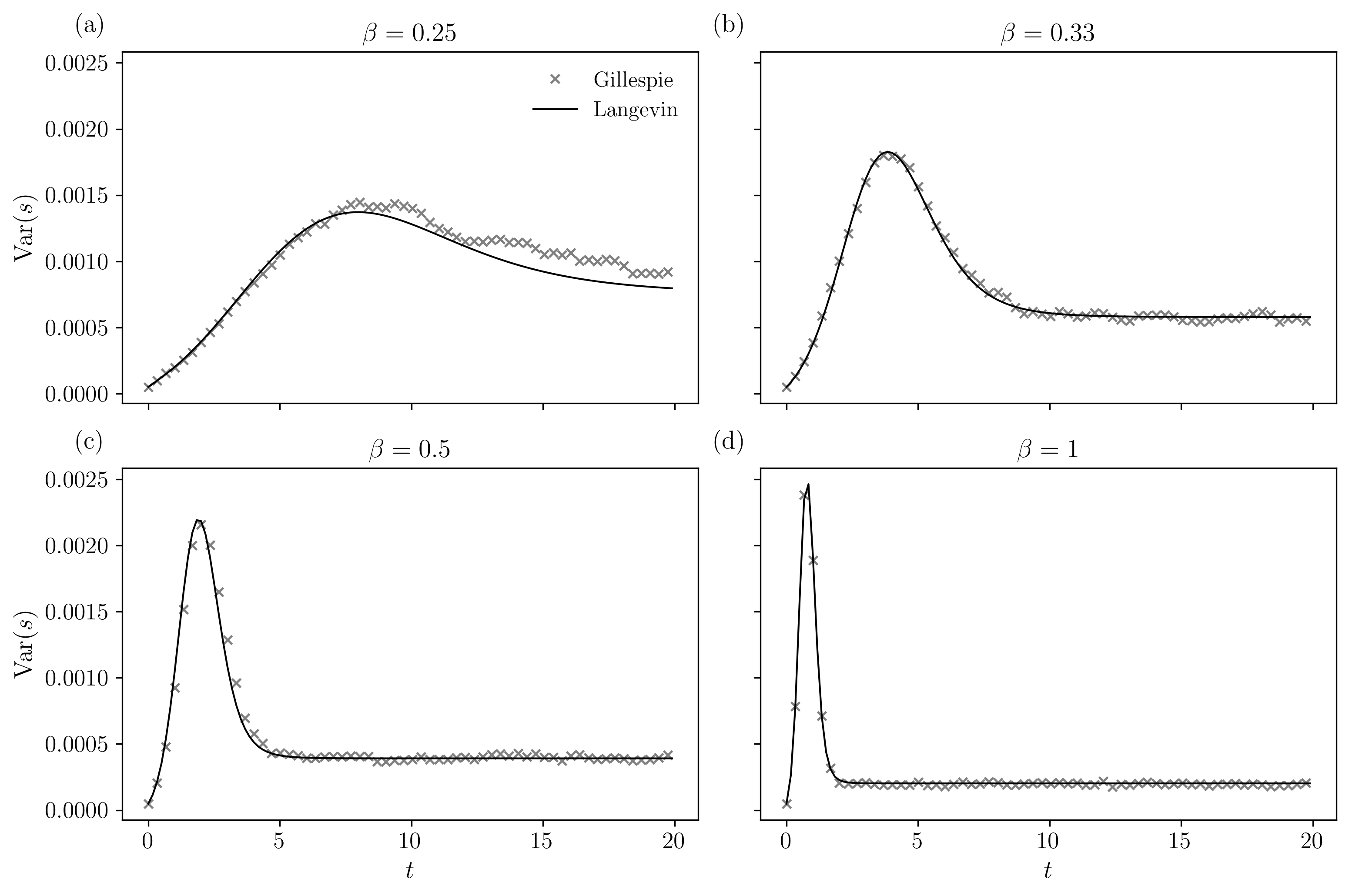}
    \caption{Comparison of the AME-Langevin predictions for $\Var(s)$ with the empirical variance measured across $N_r=1000$ Gillespie realizations of the $SIS$ process as a function of time $t$. Simulations were performed on configuration-model networks with $N=1000$ nodes and mean degree $\avg{k}\simeq 5$, generated from a truncated Poisson degree distribution with $k_{\min}=3$, $k_{\max}=20$, and Poisson parameter $\lambda=5$. The recovery rate was fixed at $\gamma=1$, while the infection rate has been varied over $\beta=0.25, 0.33, 0.5,$ and $1$. The AME--Langevin approximation reproduces both the transient peak and the long-time variance well across the parameter range, with the largest discrepancy occurring for $\beta=0.25$, where the simulated variance remains slightly above the theoretical prediction after the peak.}
    \label{fig:comparison_across_infection_rates}
\end{figure}

As previously mentioned, the Langevin construction presented above is only approximate for an $SIS$ process, as recovery reintroduces previously infected nodes into the susceptible class together with history-dependent neighborhood correlations. Consequently, while Eq.~\eqref{eq: P(m|S,k)} gives an exact conditional distribution for the neighborhood of a susceptible node in the dynamically revealed $SI$ process, it constitutes a closure approximation in the $SIS$ setting. It is therefore important to validate our predictions against direct stochastic simulations across a range of scenarios.

In all the presented simulations, the initial infected set was generated by independently infecting each node with probability $ p_0 = 0.05$. The initial covariance $C(t=0)$ used in the Langevin equation was then computed empirically from these initial conditions. Specifically, for each realization $r$ we constructed $x_r(0)=(X_{S}^{(r)}(0), X_{SI}^{(r)}(0), X_{SS}^{(r)}(0))/N$ and set $C(0)$ equal to the sample covariance, across realizations, of the fluctuation vectors $\xi_r(0) = \sqrt{N}(x_r(0)-\bar{x}(0))$, where $\bar{x}(0)$ denotes the empirical mean of $x_r(0)$ over realizations. In all simulations we present the predicted and simulated variance $\Var(s)$ of the susceptible fraction $s=X_S/N$.

We first assess the accuracy of the AME--Langevin approximation on configuration-model networks over a range of infection rates. Since our main interest lies in parameter regimes that support sustained epidemic growth, we considered values $\beta$ ranging from just above the homogeneous mean-field (HMF) estimate $\beta_c=\gamma/\langle k\rangle\simeq 0.2$ to a strongly infectious regime in which infection occurs much more rapidly than recovery.

Figure~\ref{fig:comparison_across_infection_rates} compares the predicted variance of the susceptible fraction, obtained by numerically integrating the Langevin equation~\eqref{eq:lyapunov-equation}, with the empirical variance measured across $N_r=1000$ Gillespie realizations. For each value of $\beta$, simulations were performed on a configuration-model network with $N=1000$ nodes generated from a truncated Poisson degree distribution with $k_{\min}=3$, $k_{\max}=20$, and Poisson parameter $\lambda = 5$. The recovery rate was fixed at $\gamma=1$. 

Overall, our AME--Langevin approximation reproduces both the transient peak and the long-time variance well across the full range of infection rates. The agreement is particularly close for moderate and large $\beta$, whereas near the epidemic threshold, as shown in panel~(a), the theory slightly underestimates the simulated variance, especially after the peak. This discrepancy is consistent with finite-size fluctuations near the threshold, where some infection chains may die out rapidly, while others persist for considerably longer.

\begin{figure}[!t]
    \centering
    \includegraphics[width=1.0\linewidth]{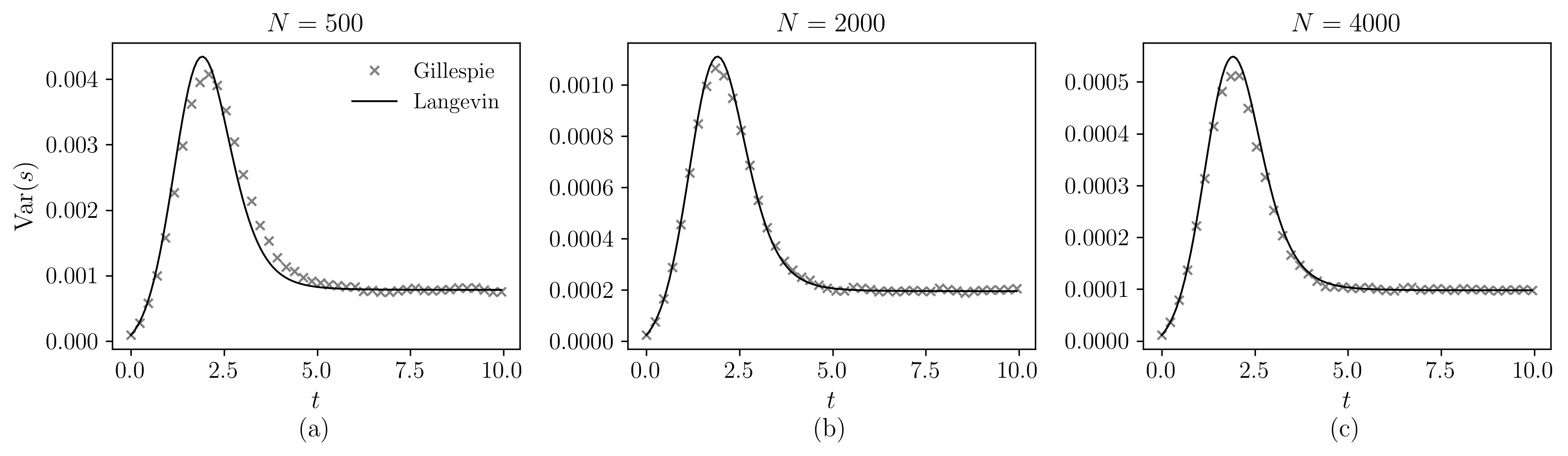}
    \caption{Comparison of the AME--Langevin predictions for $\Var(s)$ with the empirical variance measured across $N_r = 2000$ Gillespie realizations of the $SIS$ process on configuration-model networks with a truncated Poisson degree distribution, with $k_{min} = 3$, $k_{max} = 5$, and Poisson parameter $\lambda=5$. Results are shown for network sizes $N=500$ in panel (a), $N=2000$ in panel (b), and $N=4000$ in panel (c). In all cases, the infection and recovery rates were set to $\beta=0.5$ and $\gamma=1.0$, respectively. Crosses denote the Gillespie simulations, while solid lines show the AME--Langevin predictions.}
    \label{fig:sis_variance_size_sweep}
\end{figure}

We also examine the performance of the AME--Langevin approximation across networks of different sizes. Figure~\ref{fig:sis_variance_size_sweep} compares the predicted variance, with the empirical variance measured across $N_r=2000$ Gillespie realizations for $N=500$, $2000$, and $4000$, while keeping the degree distribution and dynamical parameters fixed. Once again, the approximation captures both the transient variance peak and the subsequent stationary value well for all three system sizes. The overall variance scale decreases approximately as $N^{-1}$, as expected for fluctuations of the $s$. The absolute agreement between the AME--Langevin prediction and the simulations also improves as $N$ increases, consistent with the asymptotic nature of the linear-noise approximation.

\begin{figure}[!t]
    \centering
    \includegraphics[width=1.0\linewidth]{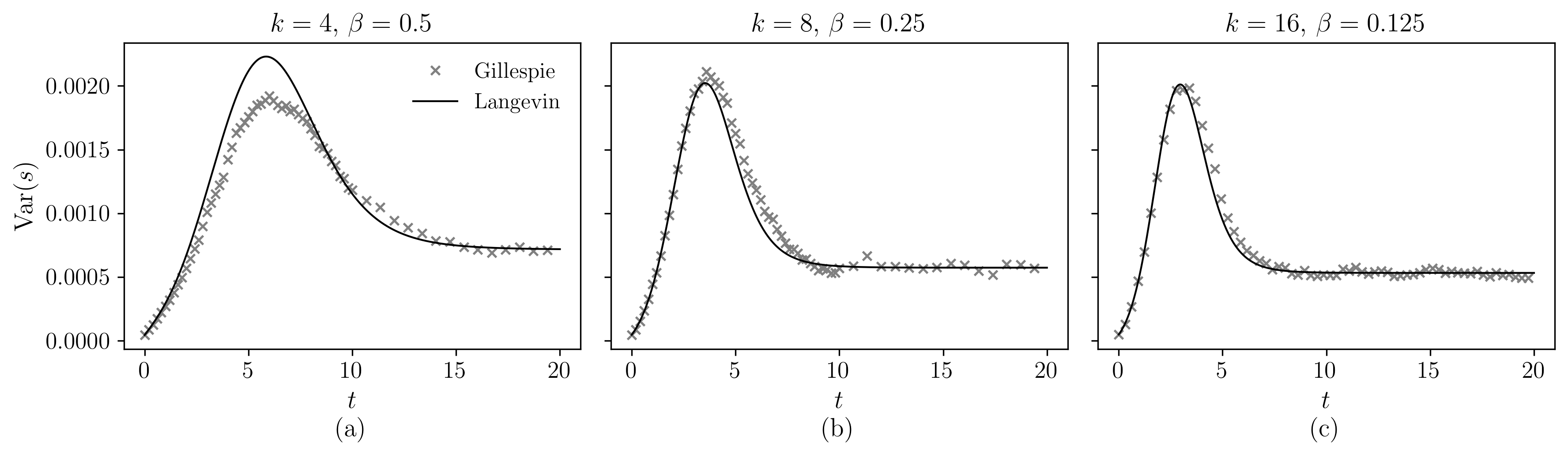}
    \caption{Comparison of the AME--Langevin predictions for $\Var(s)$ with the empirical variance measured across $N_r=1000$ Gillespie realizations of the $SIS$ process on configuration-model $k$-regular networks with $k=4,8,$ and $16$. For each value of $k$, the infection rate was chosen so that $\beta k=2\gamma$, with $\gamma=1$. All simulations were performed on networks with $N=1000$ nodes.
}
    \label{fig:k_regular}
\end{figure}

We next investigate the performance of the AME--Langevin approximation across networks with different degree distributions. We begin with $k$-regular networks, which allow us to examine how the accuracy of our approximation changes with network connectivity and sparsity in a controlled setting where the effects of degree heterogeneity are absent.

\begin{figure}[!t]
    \centering
    \includegraphics[width=0.6\linewidth]{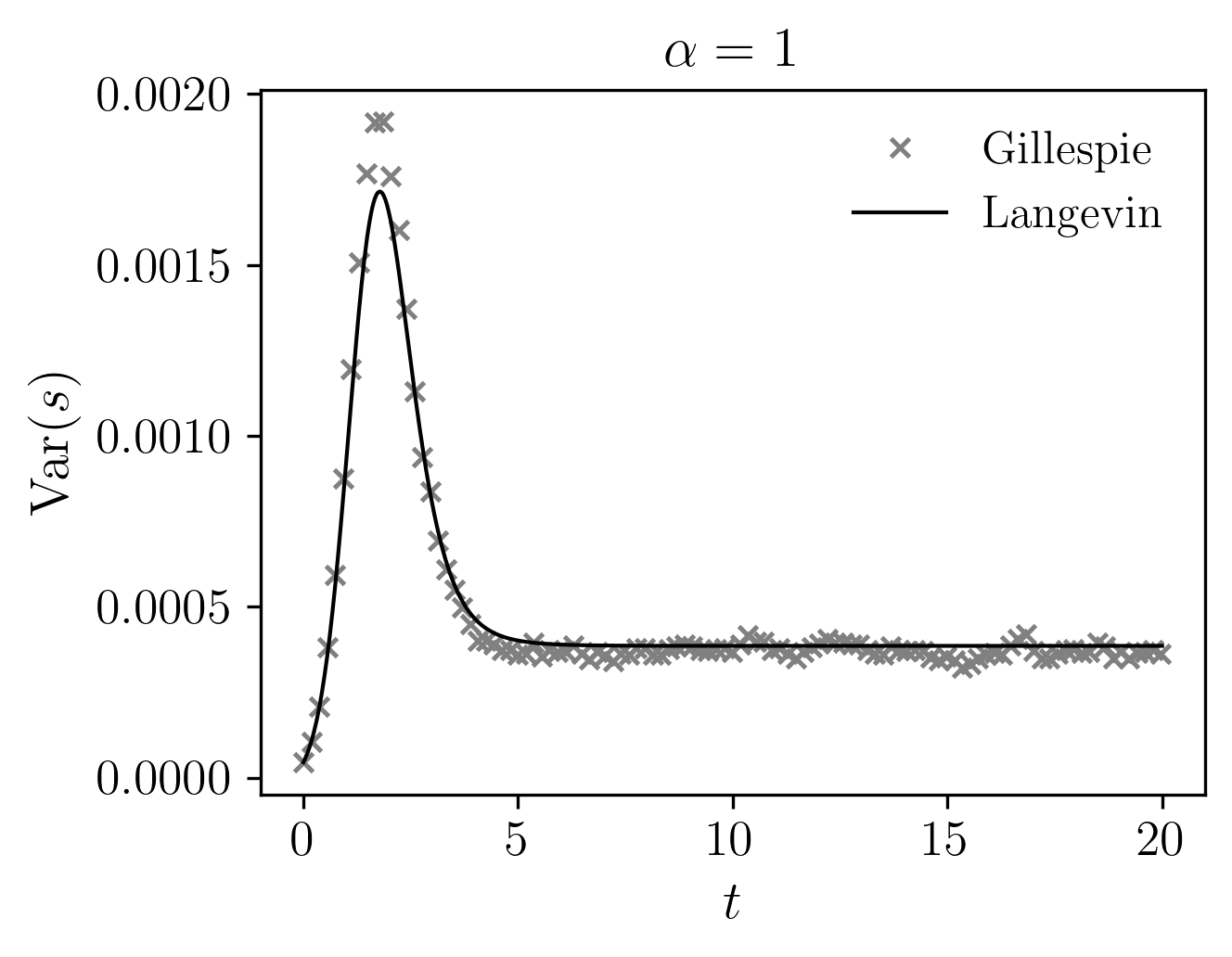}
    \caption{Comparison of the AME--Langevin prediction for $\Var (s)$ with the empirical variance measured across $N_r=1000$ Gillespie realizations of the $SIS$ process on a configuration-model network with $N=1000$ vertices and a truncated power-law degree distribution $P(k)\propto k^{-\alpha}$, with $\alpha=1$, $k_{\min}=3$, and $k_{\max}=20$. The recovery and infection rates were fixed at $\gamma=1$ and $\beta=0.25$, respectively.}
    \label{fig:sis_variance_power_law}
\end{figure}

Figure~\ref{fig:k_regular} compares the predicted variance obtained from the AME--Langevin equations with the empirical variance measured across $N_r=1000$ Gillespie realisations on configuration-model $k$-regular networks. For each $k$, the infection rate was chosen so that $\beta k = 2\gamma$, with $\gamma=1$. Since the HMF threshold for a $k$-regular network is $\beta_c = \gamma/k$, this places all three systems at the same relative distance from the threshold, namely $\beta/\beta_c = 2$. We observe good agreement between the theoretical predictions and the simulations in all cases, with the correspondence improving as $k$ increases. The largest discrepancy occurs for $k=4$, where the theory overestimates the height of the transient variance peak. A possible explanation for this overestimation is that the neighborhood-sampling closure of $P(m|S,k)$ underlying the reduced Langevin system becomes less accurate in sparse, low-degree networks. When $k$ is small, the state of each individual neighbor constitutes a substantial fraction of a node's local environment, and repeated infection and recovery events may generate strong history-dependent correlations along particular edges. These correlations may not be well represented by an approximation that samples neighborhood compositions from the global half-edge pool. As $k$ increases, each individual neighbor contributes a smaller fraction of the local environment and may therefore have less influence on the total infection pressure experienced by the focal vertex. This trend should, however, be interpreted with caution. Under the constraint $\beta k=2\gamma$, increasing $k$ is necessarily accompanied by a decrease in the per-edge transmission rate, $\beta=2\gamma/k$. Correlations associated with individual edges may consequently have a smaller effect on the total infection rate simply due to the reduced infection rate, making the aggregate neighborhood closure appear more accurate. At present, our comparison does not isolate the effect of degree alone from that of the accompanying reduction in the per-edge infection rate. We reiterate that $\kappa^{(2)}_S$ is a constant for $k$-regular graphs, since all vertices have the same degree. Consequently,
\begin{equation}
\frac{\partial \kappa_S^{(2)}}{\partial \phi(S)}=0,
\end{equation}
so the approximation introduced in Section~\ref{sec:nabla-kappa} is exact in this setting and does not affect the interpretation above.

Lastly, we validate the method on the heterogeneous graph, using the configuration model for scale-free degree distribution with minimal degree $k_{min}=3$ and maximal degree $k_{max} = 20$ and the power law exponent $\alpha=1$. As can be seen in fig.~\ref{fig:sis_variance_power_law}, while the location of the maximum of the variance is well obtained, its height is
not. We believe that the reason for that is the crudeness of the first approximation in the case of multiple hubs.

\section{Discussion}
\label{sec: Discussion}
In this paper we developed a reduced AME-Langevin approximation for the time-dependent variance of the stochastic $SIS$ process on configuration-model networks. The construction starts from Gleeson's approximate master equation, which provides the deterministic evolution of the degree- and neighborhood-resolved state variables, and combines it with a van Kampen system-size expansion for a reduced count vector consisting of the susceptible count, the count of $SI$ half-edges, and the count of $SS$ half-edges. This leads to a Langevin equation for the covariance matrix, in which the Jacobian describes the propagation of existing fluctuations and the diffusion matrix describes the stochasticity injected by individual infection and recovery events.

The numerical results indicate that the approximation captures well the full temporal profile of the susceptible variance for Poisson configuration-model networks over a range of infection rates. The agreement remains good for $k$-regular networks, especially as the degree increases. This is consistent with the intuition that the neighborhood-sampling closure becomes more accurate when individual neighbors represent a smaller fraction of the local environment. In sparse regular graphs, and especially close to the epidemic threshold, finite-size effects and repeated reinfection along the same edges become more important, and the approximation is correspondingly less accurate.

The strongest deviations occur for highly heterogeneous degree distributions. In the power-law example, the approximation captures the approximate timing of the variance peak, but not its height. This suggests that the main limitation is not the Langevin construction itself but the closure we used to approximate it. In networks with hubs, infection and recovery events repeatedly revisit a small number of structurally important nodes, generating correlations that are not well represented by sampling neighborhood states from a global pool of half-edges.

From an epidemiological perspective, the method gives access to information that is not contained in deterministic epidemic curves. The variance determines the uncertainty of prevalence forecasts, the reliability of threshold-based predictions, and the expected magnitude of stochastic deviations from the mean trajectory. This is particularly relevant in finite populations, where two realizations  can produce substantially different epidemic trajectories. A further practical advantage of the AME–Langevin approach is computational cost: solving the $3\times 3$ Langevin ODE system takes a small fraction of the time required to generate the thousands of Gillespie realizations
needed for an equivalent empirical estimate of the variance, making the method attractive for exploratory scans over parameter space or degree distributions. A computationally tractable approximation to the full time-dependent variance therefore provides a useful intermediate description between deterministic mean-field theory and large ensembles of direct stochastic simulations.\newline
 Graham and House \cite{Graham2014} and Ball et al. \cite{BallHouse2017} emphasize that, for early-growth $SIR$ dynamics on configuration-model networks, the variance depends on the third moment of the degree distribution -- one order higher than the mean -- and the AME-Langevin framework developed here makes the corresponding higher-moment dependence explicit and computable for the $SIS$ process across its full time course. In particular, the third-moment contribution, conditioned on the node being susceptible, enters through the structural factor $\kappa^{(3)}_S$, which contributes to the diffusion matrix $\boldsymbol{B}$.\newline
Several extensions of this work are natural. A first direction is to develop a more systematic treatment of the neglected derivatives of the susceptible-centered wedge factor $\kappa_S^{(2)}$, and to determine when these terms should or should not be included in the Jacobian. We believe this would be solved by keeping the Langevin system 
degree-resolved. As a next logical step, retaining a full neighborhood-resolved Langevin would eliminate the need for closures of $P(m|S,k)$. This should offer a much better framework for heterogeneous networks. Furthermore, the same framework could be adapted to other binary-state dynamics on networks, including $SIR$, $SIRS$, adaptive epidemic models, and non-epidemic spreading processes in which fluctuations around the deterministic trajectory are important. We also aim to use this tractable variance approximation to study how unknown structural information affects the quality of prediction of epidemics in networks.

\section{Acknowledgments}

\subsection*{Funding}

L.N.F., S.M.B., H.\v{S}., and V.Z. acknowledge that this work was supported by the Croatian Science Foundation
(Hrvatska zaklada za znanost) under project IP-2022-10-1648.

\subsection*{Competing interests}

The authors declare that they have no competing interests.

\subsection*{Data availability}

The data generated and analysed during the current study are available from
the corresponding author upon reasonable request.

\subsection*{Code availability}

The code used to generate the numerical results is available from the
corresponding author upon reasonable request.
\bibliography{cite}

\end{document}